# Compton wavelength, Bohr radius, Balmer's formula and g-factors

Raji Heyrovska

J. Heyrovský Institute of Physical Chemistry, Academy of Sciences of the Czech Republic, Dolejškova 3, 182 23 Prague 8, Czech Republic.

Raji.Heyrovska@jh-inst.cas.cz

**Abstract.** The Balmer formula for the spectrum of atomic hydrogen is shown to be analogous to that in Compton effect and is written in terms of the difference between the absorbed and emitted wavelengths. The g-factors come into play when the atom is subjected to disturbances (like changes in the magnetic and electric fields), and the electron and proton get displaced from their fixed positions giving rise to Zeeman effect, Stark effect, etc.

The Bohr radius ($a_B$) of the ground state of a hydrogen atom, the ionization energy ($E_H$) and the Compton wavelengths, $\lambda_{C,e}$ (= $h/m_e c$ = $2\pi r_e$) and $\lambda_{C,p}$ (= $h/m_p c$ = $2\pi r_p$) of the electron and proton respectively, (see [1] for an introduction and literature), are related by the following equations,

$$E_H = (1/2)(hc/\lambda_H) = (1/2)(e^2/\kappa)/a_B \qquad (1)$$

$$(\lambda_{C,e} + \lambda_{C,p}) = \alpha 2\pi a_B = \alpha^2 \lambda_H = \alpha^2/2R_H \qquad (2)$$

$$(\lambda_{C,e} + \lambda_{C,p}) = (\lambda_{out} - \lambda_{in})_{C,e} + (\lambda_{out} - \lambda_{in})_{C,p} \qquad (3)$$

$$\alpha = v_\omega/c = (r_e + r_p)/a_B = 2\pi a_B/\lambda_H \qquad (4)$$

$$a_B = (\alpha\lambda_H/2\pi) = c(r_e + r_p)/v_\omega = c(\tau_e + \tau_p) = c\tau_B \qquad (5)$$

where $\lambda_H$ is the wavelength of the ionizing radiation, $\kappa = 4\pi\varepsilon_o$, $\varepsilon_o$ is the electrical permittivity of vacuum, h ($= 2\pi\hbar = e^2/2\varepsilon_o\alpha c$) is the Planck constant, $\hbar$ ($= e^2/\kappa v_\omega$) is the angular momentum of spin, $\alpha$ ($= v_\omega/c$) is the fine structure constant, $v_\omega$ is the velocity of spin [2], $r_e$ ( $= \hbar/m_e c$) and $r_p$ ( $= \hbar/m_p c$) are the radii of the electron and proton [2], $m_e$ and $m_p$ are the rest masses of the electron and proton, $R_H$ is the Rydberg constant for hydrogen and $\tau_e$ and $\tau_p$ are the periods of spin of the electron and proton respectively,

$$\tau_e = 1/\omega_e = r_e/v_\omega = \lambda_{C,e}/2\pi v_\omega = \hbar/m_e v_\omega c = \lambda_{dB,e}/c \qquad (6a)$$

$$\tau_p = 1/\omega_p = r_p/v_\omega = \lambda_{C,p}/2\pi v_\omega = \hbar/m_p v_\omega c = \lambda_{dB,p}/c \qquad (6b)$$

$$\tau_B = 1/\omega_e + 1/\omega_p = (r_e + r_p)/v_\omega = a_B/c = \hbar/\mu_H v_\omega c = \lambda_{dB,H}/c \qquad (6c)$$

where $\omega_e$ and $\omega_p$ are the angular spin frequencies (and $\hbar\omega_e$ and $\hbar\omega_p$ are the corresponding energies) and $\lambda_{dB,e}$, $\lambda_{dB,p}$ and $\lambda_{dB,H}$ are the de Broglie wavelengths of electromagnetic radiation [2] associated with the spin.

Equations (2) and (3) show that $\lambda_H$, which appears in the Balmer formula [3] for the spectrum of atomic hydrogen is equal to the last term in the equation,

$$\lambda[1/(n_1)^2 - (1/n_2)^2] = \lambda_H = (1/\alpha^2)[(\lambda_{out} - \lambda_{in})_{C,e} + (\lambda_{out} - \lambda_{in})_{C,p}] \quad (7a)$$

since the Compton wavelengths of the electron and proton, are wavelength differences, between the incident ($\lambda_{in}$) and emitted (scattered) ($\lambda_{out}$) wavelengths. Thus, one can re-write the Balmer formula as,

$$\lambda[1/(n_1)^2 - (1/n_2)^2] = \lambda_H = (1/\alpha^2)[(\lambda_{out,e} + \lambda_{out,p})_C - (\lambda_{in,e} + \lambda_{in,p})_C] \quad (7b)$$

$$\lambda/(n_1)^2 = (1/\alpha^2)(\lambda_{out,e} + \lambda_{out,p})_C = x_1 \lambda_H \quad (7c)$$

$$\lambda/(n_2)^2 = (1/\alpha^2)(\lambda_{in,e} + \lambda_{in,p})_C = x_2 \lambda_H \quad (7d)$$

where $\lambda$ is the wavelength of absorbed or emitted light and $n_1$ and $n_2$ ( > $n_1$) are the principal quantum numbers. Thus $\lambda/(n_1)^2$ and $\lambda/(n_2)^2$ give important information.

Equation (5) gives a new interpretation of $a_B$ that it is the distance an electromagnetic wave travels in the period $\tau_B$, which is the sum of the periods of spin of the electron and proton.

From equations (6), the conservation laws for the linear moment of inertia, rotational (spin) moment of inertia and angular momenta are obtained, respectively, as,

$$(\hbar/c) = m_e r_e = m_p r_p = \mu_H(r_e + r_p) = \mu_H a_B \alpha \quad (8a)$$

$$(\hbar/v_\omega) = m_e r_e/\alpha = m_p r_p/\alpha = \mu_H(r_e + r_p)/\alpha = \mu_H a_B \quad (8b)$$

$$\hbar = m_e r_e c = m_p r_p c = \mu_H(r_e + r_p)c = \mu_H a_B v_\omega \quad (8c)$$

where $\mu_{red} = m_e m_p/(m_e + m_p)$ is the reduced mass of hydrogen.

The Bohr radius is also related directly to the magnetic momenta of the electron and proton (through their Compton wavelengths $\lambda_{C,e}$ and $\lambda_{C,p}$) as follows:

$$\lambda_{C,e} + \lambda_{C,p} = 2\pi(r_e + r_p) = 4\pi(\mu_B + \mu_N)/ec = 2\pi\alpha a_B \quad (9a)$$

$$(\mu_B + \mu_N) = e\hbar/2\mu_H = ev_\omega a_B/2 = ec(r_e + r_p)/2 \quad (9b)$$

$$a_B = 2(\mu_B + \mu_N)/ev_\omega = (\alpha\lambda_H/2\pi) \quad (9c)$$

where $\mu_B = e\hbar/2m_e$ and $\mu_N = e\hbar/2m_p$ are the Bohr magneton and nuclear magneton [1] respectively. The last term in equation (8c) is from equation (4) and it shows that $\lambda_H$ like $a_B$ is directly proportiona to the ratio $(\mu_B + \mu_N)/e$.

Since the magnetic moment per unit cross-sectional area ($A = \pi r^2$) is the current in a loop of circumference $2\pi r$ at a distance r from the center, the following relations hold:

$$\mu_B/A_e = i_e = (1/2)(ec/A_e)r_e = (ec/2\pi r_e) = e(m_e c^2/h) \quad (10a)$$

$$\mu_N/A_p = i_p = (1/2)(ec/A_p)r_p = (ec/2\pi r_p) = e(m_p c^2/h) \quad (10b)$$

The magnetic momenta of the free electron and proton are considered anomalous since they are both greater than $\mu_B$ and $\mu_N$,

$$\mu_e = (g_e/2)\mu_B = (g_e/2)ec(r_e/2) = (g_e/2)ev_\omega(a_{B,e}/2) \quad (11a)$$

$$\mu_p = (g_p/2)\mu_N = (g_p/2)ec(r_p/2) = (g_p/2)ev_\omega(a_{B,p}/2) \quad (11b)$$

where the g-factors (or the magnetic moment anomalies $a_e$ and $a_p$) are explained as due to the translational displacements $\delta$ during spin [2],

$$(g_e/2) = (1 + a_e) = (1 + \delta r_e/r_e) \quad (12a)$$

$$(g_p/2) = (1 + a_p) = (1 + \delta r_p/r_p) \quad (12b)$$

Since the g-factors are unity in equations (9), $\delta_e = \delta_p = 0$ and the electron and proton do not have anomalous magnetic momenta while bound in the atom. This is in agreement with the fact that $\lambda_{C,e}$, $\lambda_{C,p}$ and $a_B$ are fixed values known precisely to several decimal places [1]. The g-factors obviously come into play when the atom is subjected to disturbances (like changes in the magnetic and electric fields), and the electron and proton get free from their fixed positions giving rise to changes in $\lambda$ (Zeeman effect, Stark effect, etc.)

**Acknowledgement:** This work was financed by Grant 101/02/U111/CZ